\documentclass[sigconf,authorversion]{acmart} 

\AtBeginDocument{%
  \providecommand\BibTeX{{%
    \normalfont B\kern-0.5em{\scshape i\kern-0.25em b}\kern-0.8em\TeX}}}

\setcopyright{rightsretained}
\copyrightyear{2022}
\acmYear{2022}
\acmConference{SIGGRAPH '22 Posters}{August 07-11, 2022}{Vancouver, BC, Canada}
\acmBooktitle{Special Interest Group on Computer Graphics and Interactive Techniques Conference Posters (SIGGRAPH '22 Posters), August 07-11, 2022}
\acmDOI{10.1145/3532719.3543254}
\acmISBN{978-1-4503-9361-4/22/08}

\begin{document}

\title{Kuchibashi: 3D-Printed Tweezers Bioinspired by the New Caledonian Crow's Beak}

\author{Takahito Murakami}
\email{takahito@digitalnature.slis.tsukuba.ac.jp}
\orcid{0000-0003-2077-9747}
\affiliation{%
  \institution{University of Tsukuba}
  \streetaddress{1-2 Kasuga}
  \city{Tsukuba}
  \country{Japan}
  \postcode{305-0821}
}

\author{Maya Grace Torii}
\email{toriparu@digitalnature.slis.tsukuba.ac.jp}
\orcid{0000-0003-4025-9212}
\affiliation{%
  \institution{University of Tsukuba}
  \streetaddress{1-2 Kasuga}
  \city{Tsukuba}
  \country{Japan}
  \postcode{305-0821}
}
\author{Xanat Vargas Meza}
\email{meza@digitalnature.slis.tsukuba.ac.jp}
\orcid{0000-0003-2581-8514}
\affiliation{%
  \institution{University of Tsukuba}
  \streetaddress{1-2 Kasuga}
  \city{Tsukuba}
  \country{Japan}
  \postcode{305-0821}
}
\author{Yoichi Ochiai}
\email{wizard@slis.tsukuba.ac.jp}
\orcid{0000-0002-4690-5724}
\affiliation{%
  \institution{University of Tsukuba}
  \streetaddress{1-2 Kasuga}
  \city{Tsukuba}
  \country{Japan}
  \postcode{305-0821}
}

\renewcommand{\shortauthors}{Murakami, et al.}


\begin{teaserfigure}
  \includegraphics[width=\textwidth]{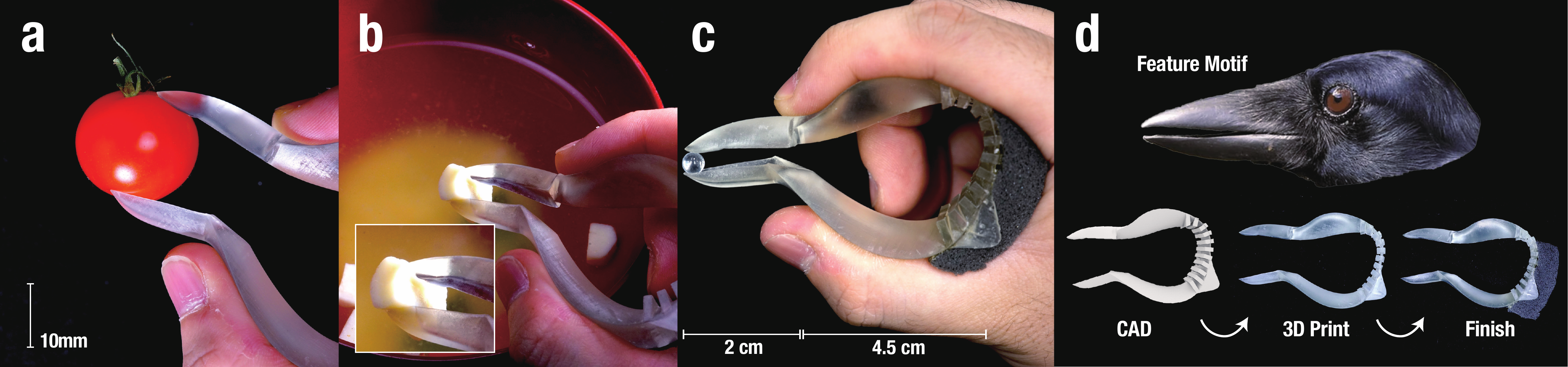}
  \caption{(a) Cherry tomato picked up, (b) tofu picked up from miso soup, (c) Kuchibashi size scale with holding hands, (d) New Caledonian crow and how to implement Kuchibashi. New Caledonian crow image from ~\cite{matsui2016adaptive} used under CC BY; background removed from the original.}
  \label{fig:teaser}
\end{teaserfigure}

\maketitle
\section{Introduction}
Humans have invented and operated tools to overcome the limitations of their physical abilities. Among these tools, tweezers are used in daily life, including in medical, engineering, and various other activities~\cite{kirkup1996history}. Despite its casual use, optimal shapes for tweezers have not received extensive research attention~\cite{kolle2016bionic}. In addition, as medical instruments, tweezers can cause anxiety in patients because of their inorganic and sharp shape ~\cite{cayer2014design}.
The aim of this study was to construct a new design for tweezers that further improves their graspability and psychological friendliness. Thus, we propose a new biomimetic gripper following the New Caledonian crow’s (NCC’s) beak shape. The NCC beak was chosen because of its superior manipulability when used as a tool and its moderately rounded and organic shape, which makes the gripper more accessible and friendlier than conventional tweezers. 

The contributions of this study are the implementation of the prototype (Kuchibashi) and user studies to evaluate the suitability of NCC beak tweezers are for human use. First, a prototype was produced and implemented to fit human hands as tweezers; it was modeled in 3D CAD and printed with a 3D printer. Second, quantitative task evaluation and qualitative surveys were conducted as user studies to compare the availability of the prototype with the conventional methods such as pinching with fingers and tweezers.

\begin{figure*}[t]
  \centering
  \includegraphics[width=\linewidth]{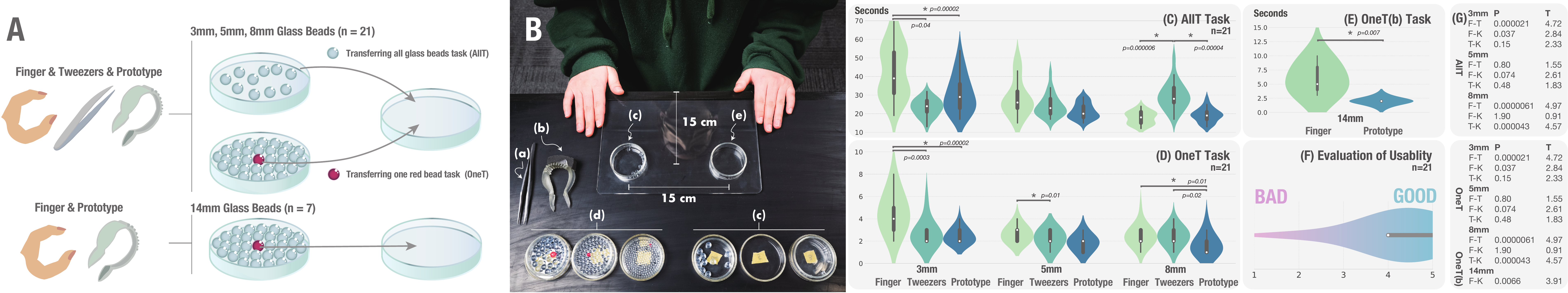}
  \caption{(A) User study design and tool combinations, and (B) setup of the experiment. Results of the (C) transferring all glass beads task (AllT), (D) transferring one red bead task (OneT), and (E) additional 14-mm task (OneT(b)). (F) Overall usability evaluation. * p<0.05. (G) P value of the user study,F:Finger, T:Tweezers, K:Kuchibashi.}
  \label{tume}
\end{figure*}
\section{Design}
Matsui et al.~\cite{matsui2016adaptive} analyzed comparing the characteristics of the NCC beak and genus \textit{Corvus}. The beak features a length of 4.3 cm from the anterior nasal hinge to the tip of the upper mandible, and the upper mandible is longer than the lower mandible (Fig.~\ref{fig:teaser}d). 

We implemented the Kuchibashi that reproduced the features of the NCC beak using 3D CAD modeling and design and 3D printing, \footnote{Rhinoceros version 6 and Formlabs Form3, Clear Resin V4 then Form Cure}, as shown in Fig~\ref{fig:teaser}d.
As shown in Fig~\ref{fig:teaser}d, the Kuchibashi features 1) a length of approximately 2 cm from the fingertip to the tip, and 2) a handgrip section of 4.5 cm. The Kuchibashi was designed for the fingers to grip approximately half the length of the beak.
\section{User Study}


In the user study, fingers, conventional tweezers, and the Kuchibashi were compared in a pinching task as a quantitative study, along with a questionnaire as a qualitative study. The pinching task was chosen to evaluate the task completion time of transferring glass beads from one container to another. The questionnaire was used to evaluate the tool's overall usability on a five-point scale with the reason for the rating as a descriptive answer, and the inclination to use the Kuchibashi in the future.

As shown in Fig~\ref{tume}A,B, the pinching experiment consisted of two tasks: transferring all 10 glass beads between petri dishes 15 cm apart (AllT), and transferring one red bead in the center (OneT). Participants performed the AllT task followed by OneT with random combinations of tool and glass bead size. The OneT task of 14-mm glass beads (OneT(b)) was performed as an additional experiment considering the results of the previous two pinching tasks.

\section{Result and Discussion}

The results of pinching task completion time were analyzed by the Shapiro-Wilk test, Kruskal-Wallis test, Steel-Dwass test, and Bonferroni correction (Fig.~\ref{tume}C,D,E). Both the AllT task and OneT task with 8-mm beads and the additional OneT experiment with 14-mm beads showed a significant difference between conventional tweezers and the Kuchibashi. 
These results show that it not only has comparable task completion time with tweezers for 3-mm and 5-mm beads but also has faster completion time for lager beads (8 mm, 14 mm).

Figure~\ref{tume}F shows the results of the five-level Likert scale overall evaluation of the Kuchibashi (mean = 4.19, SD = 0.87). There were both positive and negative opinions about the reasons for the evaluation. Positive opinions included
"It was easy to grab because it was hard to slip," "There was a sense of security in being caught,"
"I felt secure being caught," and 
"The design was fashionable."
Negative opinions included
"Too thick to cope with small items."
There were also descriptions of other use senarios, such as "I want to use it as a tableware."
Eighty percent of the participants answered "yes" to whether they would like to use Kuchibashi in the future. 
From these survey results, the Kuchibashi design, especially its impression of security and safety, was perceived positively overall by the participants.

However, the design has several limitations. First as mentioned in the pinching task result, the Kuchibashi is not significantly quicker than the other method, such as using finger or conventional tweezers, when the beads are small. In addition, with the current shape and size of Kuchibashi, extremely small objects are difficult to manipulate. Furthermore, there are limitations in our user study in that the task was not performed with shapes, material or sizes other than 3-mm to 14-mm beads. In addition, a comparison with other pinching tools, such as chopsticks and tongs, was not considered and compared in this user study.

\section{Conclusion and Future Work}
In this study we have implemented the Kuchibashi, an NCC beak-like tweezers, and conducted a user study to evaluate the Kuchibashi's usability. 
We proved through experiments that the proposed Kuchibashi design is functionally not inferior to other major pinching methods in interacting with small objects (T = 0.43, SD = 0.60) and is superior in interacting with larger objects (T = 3.14, SD = 0.86).
As future research, the tool's use as tableware can be explored considering Kuchibashi's advantages; thus, a new Kuchibashi and user study can be designed.




\begin{acks}
This work was supported by ARE from the University of Tsukuba.
\end{acks}

\bibliographystyle{ACM-Reference-Format}
\bibliography{reference}


\end{document}